\documentclass[12pt,preprint,epsf]{revtex4}
\usepackage{epsfig}
\usepackage{graphics}% Include figure files
\usepackage{bm}% bold math
\begin{document}
\title{
%%\bf The generalized TRK-GGT dispersion sum rules for total photoabsorption
%%cross sections on nucleons and few-body nuclei
Dispersion total photoproduction sum rules for nucleons and few-body nuclei revisited
%%Continuum spectroscopy of Borromean two-neutron halo nuclei
}
\author{S.B.~Gerasimov}
\email{gerasb@theor.jinr.ru}
\affiliation
{\it Joint Institute for Nuclear Research,141980 Dubna, Russia}
%%\thanks{Permanent address:
%%{Joint Institute for Nuclear Research, 141980 Dubna, Russia},
%%\author{B.V.~Danilin}\thanks{Permanent
%%address: Russian Research Center "The Kurchatov Institute", 123182
%%Moscow, Russia}
%%\author{J.S.~Vaagen}\affiliation{Department of Physics and
%%Technology, University of Bergen, Norway}
%%\date{empty}
%\begin{document}
%\renewcommand{\baselinestretch}{1.0}
\begin{abstract}
The questions on the presense and quantitative role of the
constant terms in the real part of the high-energy photon-nucleon
and photon-nucleus amplitudes representing the contribution of the
non-Regge (the fixed $j=0$-pole) singularities in the
finite-energy sum rules (FESR) for the photoabsorption cross
sections on nucleons and the lightest atomic nuclei are discussed
and new testable relations are presented for relevant combinations
of the Compton scattering amplitudes.
%Energy- and angular correlation distributions of the three
%fragments in $^6$He breakup on $^{208}$Pb at collision energy 240
%MeV/nucleon are discussed within the microscopic four-body
%distorted wave model and compared with experimental data. The
%nuclear structure of the ground state and low-energy three-body
%continuum of $^6$He is calculated by the method of hyperspherical
%harmonics within the three-body cluster model. Reflections of the
%fundamental permutation symmetry of the halo neutrons in angular
%and energy correlations are pointed out. The calculations describe
%the experimental data for fragment correlations near breakup
%threshold rather well and the physics is contained in a few
%elementary modes, but with increasing excitation energy of $^6$He
%some striking deviations from experimental distributions are
%encountered. Possible reasons for this are discussed.
\end{abstract}
%%\pacs
%%{13.60.Fz, 12.40.Nn, 25.20.Dc, 11.55.Hx}
%%21.45.+v, 21.60.Gx, 25.60.-t, 25.75.Gz
\maketitle

\section{Introduction}
 %Text of section.
In the 1954 seminal paper of Gell-Mann, Goldberger and Thirring
(GGT)\cite{GGT} on the use of the causality condition in quantum
theory the idea of the "superconvergence" sum rule technique was
first suggested and applied to the photonuclear absorption
processes. The GGT sum rule follows from the assumption of
validity of the unsubtracted dispersion relations for the
difference presumably vanishing as $\nu \to \infty$
\begin{eqnarray}
\Delta T=T_{\gamma A}(\nu)-ZT_{\gamma p}(\nu)-NT_{\gamma n}(\nu)
\end{eqnarray}
of the forward Compton scattering amplitudes on the nucleus with
atomic number $A=Z+N$ and the sum of amplitudes on the Z free
protons and N free neutrons.~~After inclusion of the Thompson
value $-(\alpha Q^2)/M$ (M(Q) being the hadron mass and electric
charge in the units of the electron charge), for every hadron
amplitude $T(\nu=0)$ at zero photon energy, the sum rule reads
\begin{eqnarray}
2\pi^2\frac{\alpha}{M_{n}}(\frac{-Z^2}{A}+Z)+
\int_{\nu_{\gamma\pi}}^{\infty}d\nu [Z\sigma_{\gamma p}(\nu)
+ N\sigma_{\gamma n}(\nu)
-\sigma_{\gamma A}(\nu)]
= \int_{\nu_{thr}}^{\nu_{\gamma\pi}} d\nu \sigma_{\gamma A}(\nu)
\label{GGT}
\end{eqnarray}
The first term in l.h.s. of (\ref{GGT}) practically coincides with
the "kinetic" part of the long-known Thomas-Reiche-Kuhn sum rule
for the electric dipole nuclear photoabsorption
\begin{eqnarray}
\sigma_{0}(E1) &\equiv& \int_{\nu_{thr}}^{\infty}\sigma_{E1}(\nu)d\nu=
4\pi^2 \Sigma_{n} (E_{n}-E_{0})|\langle n|D_{z}|A\rangle|^2
=2\pi^2 \langle A|[D_{z}[H,D_{z}]]|A\rangle \nonumber \\
&=&(2\pi^2\alpha NZ)/(AM_{n})+2\pi^2 \langle A|[D_{z}[\hat{V}_{NN},D_{z}]]|A\rangle
\end{eqnarray}
where the first term results from the double commutator with the
kinetic energy operator of the nuclear hamiltonian $H$. The
present work originates partly from earlier papers of the author
\cite{Ge63,Ge64,Ge72} dealing with sum rules for total
photon-hadron cross-sections and aims to present some new
experimentally testable and theoretically interesting relations
emerging from the dispersion FESR phenomenology.
\section{Towards generalized GGT sum rule}
It was always tempting and rewarding to combine the power of
dispersion relation approach, which is based on very general
underlying assumptions and explains many general properties of the
scattering amplitudes as well as provides useful relations between
them in a rather simple way with particular dynamical ingredients
of a given quantum system such as, for instance, implications of
the broken chiral symmetry and pionic dynamics dominating
peripheral properties and low-energy interactions of hadrons and
nuclei.
\subsection{A glance at a possible role of pion degrees of freedom on GGT sum rule}
In \cite{Ge63,Ge72} an attempt was maid to introduce corrections
to the GGT approach understood as a familiar Impulse Approximation
(IA) scheme applied to the $\gamma A$~-forward scattering
amplitude. The approximate relevance of IA is seen from the fact
that it corresponds to taking into account the singularities
closest to the physical region of the peripheral scattering
process ($t\leq 0$, $t=(k-k^{'})^2$ is the invariant 4-momentum
transfer for elastic scattering). The respective cut in the
complex $t$-plane is defined by the diagrams schematically
represented in Fig.1a, while the next to the leading "anomalous"
threshold given by Fig.1a will be the "normal" $2\pi$~-exchange
diagrams, represented in Fig.1b, with the cut starting at
$t=4m_{\pi}^2$.
\begin{center}
Fig.1
\end{center}
\hspace*{.5cm}~~$\sim\sim\sim \bigcirc \sim\sim\sim$~~~~~~~\hspace*{1.3cm}$\sim\sim\sim\bigcirc\sim\sim\sim$\\
\hspace*{1.5cm}$\nearrow \searrow$ \hspace*{4.23cm} $\vdots \  \vdots$ \\
(a)~$\Longrightarrow \bigcirc \longrightarrow\bigcirc \Longrightarrow$~~~
\hspace*{.75cm}(b)~~$\Longrightarrow \bigcirc \Longrightarrow $ \\
In Fig.1, the solid lines refer to nucleons and nuclei, the wavy
(dotted) lines represent photons and pions. Graph $(a)$ represents
the impulse approximation (IA), while $(b)$ defines the correction
related with the nuclear "collective" pion cloud and thus is
effective due to short-ranged $NN$-correlation inside nuclei.
Their relative role can qualitatively be characterized by the
ratio
\begin{eqnarray}
 \frac{t_{0}(IA)}{t_{0}(2\pi)}\simeq \frac{8m_{n}\varepsilon_{b}}
{(A-1)\cdot4 m_{\pi}^{2}}
\label{t0}
\end{eqnarray}
where $t_{0}$ refers to the beginning of the respective cut in the
complex $t$-plane, $m_{n}$ is the nucleon mass, $\varepsilon_{b}$~
is the nuclear binding energy. For instance, this ratio is $\sim$
.22 (.40 and .66) for the $d$ $(^{3}He$ and $^{4}He$,
respectively). This indicates that, naturally, for~~$^{3}He$ and
$^{4}He$ the "pionic" contributions will be significantly more
important compared to deuteron. Equation (\ref{t0}) also signals
that, in the considered respect, the situation for heavier nuclei
is expected to be much alike the $^{4}He$ case because of nearly
equal binding energy per nucleon.
\subsection{Towards the measurement and systematization of $<A|\phi^{*}\phi|A>$}
A further step in relevant implementation of pionic d.o.f. into
the GGT sum rule was an observation inferred from models providing
the convergence of the $\sigma_{0}(tot)$-integral. It was first
suggested \cite{Ge64} and then perturbatively (to one-loop order)
checked \cite{GeMo75} in scalar, $\phi^3$-type
"super-renormalizable" model that the generalized
Thomas-Reiche-Kuhn is valid for total photoabsorption cross
section
\begin{eqnarray}
 \sigma_{0}=\int d\nu \sigma_{tot}(\nu)=2\pi^2\langle
 \phi_1|[D[H,D]]|\phi_1\rangle,
\end{eqnarray}
where the charged scalar field $\phi_1$ is locally connected with
two scalar fields, $\phi_2$ being charged one and the other,
$\phi_3$,neutral. The double commutator is then interpreted via
the known Schwinger-term, {\it i.e.}, the equal-time commutator of
the time- and spatial-component of e.m. current operator. Hence,
the generalized, "GGT$^{'}$"-sum rule, implicitly including the
integrals of the absorptive parts of the amplitudes presented by
the diagrams with $2\pi$-exchanges,was written \cite{Ge72} in the
form
\begin{eqnarray}
\sigma_{0}^{\gamma A}-Z\sigma_{0}^{\gamma p}-N\sigma_{0}^{\gamma
n}= 2\pi^2\alpha[\frac{NZ}{Am}+\int d{\vec x}(\langle
A|\phi^{*}\phi|A\rangle -\Sigma_{i}\langle
N_i|\phi^{*}(x)\phi(x)|N_i\rangle)].
\end{eqnarray}
The photonuclear sum rule including the terms
$<A|\phi^{*}\phi|A>$~~and $<N|\phi^{*}\phi|N>$, represented by the
Feynman diagram in Fig.2
\begin{center}
Fig.2
\end{center}
\begin{center}
$\sim\sim\sim\bullet\sim\sim\sim$ \\
$\vdots \ \vdots$ \\
$\Longrightarrow \bigcirc \Longrightarrow$
\end{center}
was later rediscovered \cite{Ku91}, found to be a useful
exploration tool \cite{MEr98} and widely discussed ({\it e.g.},
\cite{MEr00} and further references therein) in view of the
interesting idea about possible partial restoration of the chiral
symmetry in real nuclei.

The matter is that up to constant factors
the matrix element corresponding to the seagull graph in Fig.2,
which in the forward direction is gauge invariant and may have a direct
bearing to measurable quantities, is essentially of the same structure as
the matrix element represented by Fig.3
\begin{center}
Fig.3
\end{center}
\begin{center}
$\otimes $ \\
$\vdots \ \vdots $  \\
$\Longrightarrow \bigcirc \Longrightarrow$
\end{center}
The symbol $\otimes$ denotes the local, scalar quark-current and therefore it
is directly connected with the $\Sigma_{\pi}$-term, hence with the chiral symmetry breaking
and its possible (partial) restoration in nuclear matter.
\subsection {FESR and problem of "Big Circle" contribution}
The standard FESR technique enables one to deal with the
amplitudes defined
in the finite region of the complex energy plane \\
\begin{eqnarray}
 f(\nu)=\frac{1}{2\pi\imath}\oint dz \frac {f(z)}{z-\nu}
\end{eqnarray}
where $f(\nu)$ is the spin-averaged, forward Compton scattering
amplitude and the integration contour includes both sides of the
cuts along the real axes $-R\leq \nu \leq R$ closed by a circle of
a "big" radius R. As usual the problem consists in the justified
and economical choice for the representation of amplitudes in the
complex energy plane to fulfill the integration over the large but
finite-radius circle in the complex plane. We keep the original
GGT idea of a relation between the photon-nucleus scattering
amplitude and a relevant combination of the photon-nucleon
amplitudes at sufficiently large photon energies, but our choice
of the "superconvergent" combination of Compton amplitudes
$f_{\gamma A(p,n)}$ is different from GGT.It includes amplitudes
of two nuclei with $A_1=Z_1+N_1,A_2=Z_2+N_2$~and is assumed to
satisfy the condition:
\begin{eqnarray}
lim_{|\nu|\to R}[\frac{1}{A_1}f_{A_1}-\frac{1}{A_2}f_{A_2}]=\frac{Z_1N_2-N_1Z_2}{A_1A_2}
(f_p-f_n)|_{\nu=R} + \frac{S_{\pi}(A_1)}{A_1}-\frac{S_{\pi}(A_2)}{A_2}
\label{asym}
\end{eqnarray}
where
\begin{eqnarray}
S_{\pi}(A_{i}) \simeq \frac{\alpha}{3} \int d^3x \langle A_{i}|\vec{\phi}(x)\vec{\phi}(x)|A_{i}\rangle
\label{sigma}
\end{eqnarray}
and the scalar product in the integrand is understood to be
in the isospin space. \\
The upper limit $\nu_{max}\equiv R$ in all integrals should be chosen
from the compromise provisions.

The first term derived in the approximation linear in
$A_{i},~(i=1,2)$~ is parameterized through the
$a_2(J^P;I^G=2^{+};1^{-})$~-Reggeon exchange in the $t$-channel
and includes in addition the real constant term seemingly taking
place \cite{DaGi70} in the $Re f_{p}$ and referring as the residue
of the $j=0$ fixed-pole in the complex angular momentum plane.
Hence one should  put $R\ge 1.5\div 2.0$~GeV to apply the
Regge-pole phenomenology with the commonly used parameters
\cite{He70}
\begin{eqnarray}
Im[f_{p}(\nu)-f_{n}(\nu)]=(\nu/4\pi)
%\frac{\nu}{4\pi}
(\sigma_{p}^{tot}-\sigma_{n}^{tot}) =b_{a_2}\nu^{1/2} \nonumber \\
Re (f_{p}(\nu)-f_{n}(\nu))=
(1/4\pi)b_{a_2}(-\nu^{1/2})+C_{p}-C_{n} \nonumber   \\
\sigma_{p}^{tot}(\nu)- \sigma_{n}^{tot}(\nu)=24.6/\nu^{1/2}
\label{isovec}
\end{eqnarray}
Following \cite{DaGi70}, we accept $C_{p}\simeq-3.0 \mu b\cdot
GeV$ and put the $C_n$-value rather arbitrarily to be either
$C_{n}=(2/3)C_p$ or
$C_{n}=0$ for the sake of further numerical estimations.\\
Due to the dominant scalar-isoscalar nature of the pionic
operators we accept $S_{\pi}(p)\simeq S_{\pi}(n)$
while
$S_{\pi}(A_{i})\neq 0$ will disclose its essential nonlinear
dependence on the atomic number $A$ of real nuclei.
\section{Numerical results and discussion}
As an example of the generalized nuclear sum rule applications, we
choose a pair of lightest nuclei - the deuteron and $^{3}He$.
While in the deuteron case the total photoabsorption cross section
is known well above our taken $\nu_{max}\simeq 1.6~GeV$, the
$\sigma_{tot}(\gamma ^{3}He)$ is known to $.8~GeV$
\cite{Daresbury}  ; hence, in this case, we have to take
$\nu_{max}=.8~GeV$. The major purpose of using these new types
nuclear sum rules may be the extraction of information about the
value of difference of the nuclear matrix elements: $\Delta
\sigma_{\pi}= \int
d{\vec{x}}\frac{m_{\pi}^2}{2}[\frac{1}{A_1}<A_1|\vec{\phi}(x)\cdot\vec{\phi}(x)|A_1>
-\frac{1}{A_2}<A_2|\vec{\phi}(x)\cdot\vec{\phi}(x)|A_2>]$.\\
The term $\Delta \sigma_{\pi}$ can thus be extracted from
experimentally measurable quantities to give useful information on
the values closely related with the chiral symmetry
characteristics in real nuclei. Of special interest is the
situation when $Z_1N_2-N_1Z_2=0$ in (\ref{asym}), as for the
deuteron- and $^{4}He$-pair, to mention. The contribution of the
$a_2$-Reggeon is then absent and the optimal value of
$\nu_{max}=R$ in dispersion integrals of cross sections could
probably be taken at a lower value. Qualitatively, this newly
chosen $R$-value should provide a reasonable balance between the
contribution of the same group of most important nucleon
resonances into the real parts of nuclear Compton amplitudes
represented by the terms $S_{\pi}(A_{i})$ and the respective
imaginary parts entering into dispersion integrals in the form of
the corresponding nuclear photo-pion production cross sections.

For arbitrary $A_1=Z_1+N_1$ and $A_2=Z_2+N_2$ our general sum rule reads
\begin{eqnarray}
2\pi^2[\frac{f_{A_{1}}(\nu=0)+S_{\pi}(A_{1})}{A_{1}}-
\frac{f_{A_{2}}(\nu=0)+S_{\pi}(A_{2})}{A_{2}} + \nonumber \\
+\frac{Z_{1}N_{2}-Z_{2}N_{1}}{A_{1}A_{2}}
\cdot(\frac{2b_{a_{2}}\nu_{max}^{1/2}}{2\pi^2}
-C_{p}+C_{n})]
=\frac{\sigma_{0}^{\nu_{max}}(\gamma A_{1})}{A_{1}}
-\frac{\sigma_{0}^{\nu_{max}}(\gamma A_{2})}{A_{2}}
\end{eqnarray}
where $f_{A_{i}}(\nu=0)\simeq -(\alpha Z_{i}^2)/(A_{i}m_{n})$ is the
Thompson zero-energy amplitude, $S_{\pi}(A_{i})$ is defined in Eq.(\ref{sigma})
and the integration in $\sigma_{0}^{\nu_{max}}$ extends from the photodisintegration
threshold to the upper bound $\nu_{max}$. In the case of $^{3}He$ and deuteron
the integration was carried out with the cross-sections tabulated
in \cite{data} up to $\nu_{max}=.8~GeV$. The low-energy integrals
up to the pion photoproduction thresholds $\nu_{\gamma \pi}$ were approximated by
\begin{eqnarray}
\sigma_{0}^{\nu_{\gamma \pi}} = 60 \frac{NZ}{A}(1+\kappa_{A}^{exp})~[\mu b\cdot GeV]
\end{eqnarray}
where $\kappa_{He-3(d)}^{exp}=.75\pm.10~(.37\pm.11)$, following~
\cite{Drech}. To have an idea about the scale of $S_{\pi}$ for the
nuclei considered, we confronted the calculated values of
$$(2\pi^2\alpha)/(3)[(1/3)S_{\pi}(^{3}He)-
(1/2)S_{\pi}(d)]\simeq 7.75~(1.17)~ [\mu b\cdot GeV]$$
for $C_p=-3,C_n=-2~(0)$~with the value
$$60\cdot[(2/9)\kappa_{^{3}He}-(1/4)\kappa_{d}]=
(1/3)\cdot 40\cdot (.75\pm.10)-(1/2)\cdot 30\cdot (.37\pm.11) \simeq
4.4\pm 2.1~[\mu b\cdot GeV]$$
representing the "potential parts" in the
difference of non-relativistic TRK sum rules
$$ 2\pi^2 \alpha[(1/3)<^{3}He|[D,[V_{NN}]]|^{3}He>-
(1/2)<d|[D,[V_{NN}]]|d>]$$ The correspondence looks reasonable
because the non-relativistic value is in between two values
following from a more general sum rule with the differing values
of $C_n$. We also draw attention to a strong dependence of the
mentioned estimations on two chosen numerical values of $C_n$,
which emphasizes the significance of sum rule as a source of new
interesting information.
\section{Conclusion}
The pion-nucleon sigma-term
%is defined as
\begin{eqnarray}
\sigma = \frac{\hat{m}}{2 m_p} \langle p | \bar{u}u+\bar{d}d|p
\rangle, \hspace{1.cm} \hat{m}=\frac{1}{2}(m_u+m_d), \nonumber
\end{eqnarray}
and, generally, sigma-terms of a given hadron are proportional to
the scalar quark currents
\begin{eqnarray}
\langle A|m_q \bar{q}q|A \rangle \; \; ; q=u, d, s \; \; ; \
A=\pi, K, N,~^{Z}A_{N}-nuclei. \nonumber
\end{eqnarray}
These are of great physical significance because they are related
to the hadron masses, to the meson scattering amplitudes
\cite{Sa01}, to the strangeness content of $A$, and to the
properties of nuclear \cite{Gi05} and dark \cite{Bo02} matter. Our
derived and discussed photoabsorption sum rules are focused on the
comparison of the scalar pion densities, hence, in part, of the
pionic $\sigma$-terms for different nuclei to trace their
dependence on the atomic number. The deuteron sum rule provides
thereupon the situation most close to free nucleons while the
helium-4 plays the role of a drop of the real nuclear matter.

In view of the above discussion the following looks to be practically important: \\
 1. To extend measurements of the total photoabsorption on
the $^{3}He$~ and $^{4}He$~-nuclei at least up to energy of photons $1.5\div 2.0~~GeV$.\\
2. To complete calculation of $<A|[D[H,D]]|A>$,~$A=^{3(4)}He$,
with best modern potentials and
respective wave functions as well as with estimation of relativistic corrections. \\

{\small Partial support of this work by the Blokhintsev-Votruba
foundation is gratefully acknowledged. }

%%\end{document}
\end{document}